\documentclass{mem}

\usepackage{natbib}
\usepackage{txfonts}
\usepackage{balance}
\usepackage{graphicx}
\usepackage{flushend}
\usepackage{xcolor}
\usepackage{aas_macros}
\usepackage{hyperref}
\idline{75}{282}
\begin{document}
\def\teff{$T\rm_{eff }$}
\def\kms{$\mathrm {km s}^{-1}$}

\title{
Investigating Active Galactic Nuclei variability with the Cherenkov Telescope Array Observatory
}

\author{
G. \,Grolleron\inst{1} \and J. \,Biteau\inst{2} \and M. \,Cerruti\inst{3} \and R. \,Grau\inst{4} \and L. \,Gréaux\inst{2} \and T. \,Hovatta\inst{5} \and J.-P. \,Lenain\inst{1} \and E. \,Lindfors\inst{5} \and W. \,Max-Moerbeck\inst{6} \and
D. \,Miceli\inst{7} \and A. \,Moralejo\inst{4} \and K. \,Nilsson\inst{5} \and E. Prandini\inst{7} \and E. \,Pueschel\inst{8} \and S. \,Kankkunen\inst{9,10} for the CTAO Consortium
\\
The remaining authors can be found at the end of the paper}

\institute{
Sorbonne Université, CNRS/IN2P3, Laboratoire de Physique Nucléaire et de Hautes Energies, LPNHE, 4
place Jussieu, 75005 Paris, France
\and 
Laboratoire de Physique des 2 infinis, Irene Joliot-Curie, IN2P3/CNRS, Université Paris-Saclay, Université de Paris,
15 rue Georges Clemenceau, 91406 Orsay, Cedex, France
\and
Université de Paris, CNRS, Astroparticule et Cosmologie,
10 rue Alice Domon et Léonie Duquet, 75013 Paris Cedex 13, France
\and
Institut de Fisica d'Altes Energies (IFAE), The Barcelona Institute of Science and Technology,
Campus UAB, 08193 Bellaterra (Barcelona), Spain
\and
Finnish Centre for Astronomy with ESO (FINCA), University of Turku,
Vesilinnantie 5, 20014 University of Turku, Finland
\email{guillaume.grolleron@lpnhe.in2p3.fr}\\
}

\authorrunning{Grolleron}

\titlerunning{AGN variability}


\abstract{
Blazars, a type of active galactic nuclei (AGN) with relativistic jets pointed at the observer, exhibit flux variability across the electromagnetic spectrum due to particle acceleration in their jets. Power spectral density (PSD) studies show breaks at specific frequencies, particularly in X-rays, linked to the accretion regime and black hole mass. However, very-high-energy gamma-ray PSD breaks remain unexplored due to current instrument limitations. The Cherenkov Telescope Array Observatory (CTAO), with up to ten times greater sensitivity compared to current generation instruments, will allow precise PSD reconstruction and unprecedented study of blazar flares. These flares reveal key insights into particle acceleration, photon production, and jet properties. The AGN monitoring and flare programs in CTAO's Key Science Project aim to deepen our understanding of blazar emissions.
\keywords{Active galactic nuclei, CTAO, AGN variability.}
}
\maketitle{}

\section{Introduction}
The Cherenkov Telescope Array Observatory (CTAO) represents a major leap in the filed of ground-based imaging atmospheric Cherenkov telescopes (IACT), improving sensitivity with respect to current instruments by a factor of 5 to 10 above 20 GeV, depending on the energy range. This will provide unprecedented access to the Universe's non-thermal emissions. The array consists of two sites: La Palma, Spain, for extragalactic sources, and Paranal, Chile, for galactic sources. Active Galactic Nuclei (AGN) are key multiwavelength emitters, driven by both thermal and non-thermal processes. Relativistic jets lauched from their central black holes accelerate particles to extreme energies, producing significant non-thermal radiation. Blazars, a subset of AGNs with jets directed towards Earth, show intense variability in flux and spectrum, ranging from minutes to years \citep{2019NewAR..8701541H}. Such variability offers insights into emission regions, mechanisms, and particle acceleration processes \citep{2019ARA&A..57..467B, 2020Galax...8...72C}.

Long-term AGN studies reveal power spectral density (PSD) transitions from pink to red noise \citep{2019Galax...7...28R}. While X-ray studies have linked PSD break frequencies to black hole mass and accretion rate \citep{1999ApJ...514..682E, 2002MNRAS.332..231U, 2006Natur.444..730M}, these breaks are hard to measure at very high energies with current instruments. CTAO’s advanced sensitivity would enable such reconstructions for more sources.This work highlights AGN flare studies and long-term monitoring in CTAO, part of the AGN Key Science Project \citep{inbook}. The project plans weekly observations of 18 AGNs over a decade to reconstruct flux distributions and duty cycles of jetted AGNs while exploring rapid jet variability and emission mechanisms.

Building on \cite{2023arXiv230414208C, 2023arXiv230912157G, 2023arXiv230909615C}, we simulate 20 years of BL Lac observations to evaluate flux lightcurve (LC) and PSD reconstruction. For AGN flares, model-based simulations of a Mrk 421 flare analyze CTAO's ability to detect spectral variability using hardness ratio diagrams. The following sections discuss AGN models, CTAO observation simulations, spectral and flux reconstruction methods, results, and prospects for CTAO's future contributions.

\section{AGN modelling}

\subsection{Modelling of the long term AGN behavior}
\begin{figure}[]
\begin{center}
\resizebox{.45\textwidth}{!}{\includegraphics[clip=true]{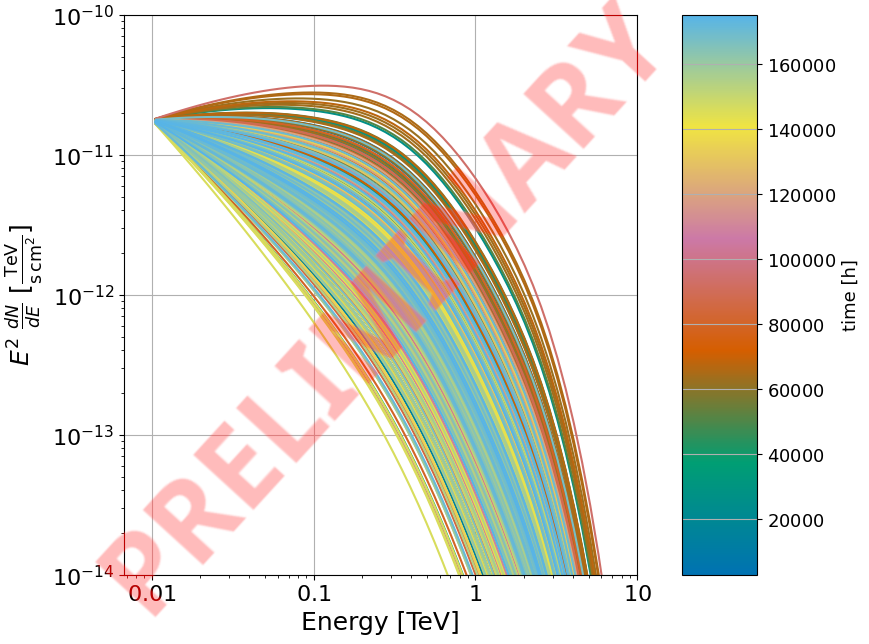}}
\end{center}
\caption{The generated spectra for BL Lac. Colors displaying the time evolution.}
\label{fig:SED_BLLac} 
\end{figure}

The AGN long-term modeling approach is detailed in \cite{2023arXiv230912157G}. Briefly, we generate time series for flux normalization and photon index using the framework from \cite{2013MNRAS.433..907E}, with the following assumptions. First, we assume that the distribution of the observed flux is log-normal. Secondly, the fractional variability ($F_\mathrm{var}$), as defined in \cite{2003MNRAS.345.1271V}, is scaled to match observations of the bright blazars Mrk 421 and Mrk 501 from current IACT data \citep{2023arXiv230400835G}. The time series for photon index follows the "harder-when-brighter" trend \citep{2010A&A...520A..83H,2014MNRAS.444.1077K}, where the very-high-energy spectrum hardens as flux increases.
Finally, the time-dependent evolution of the AGN gamma-ray spectra is modeled with $\Phi(E,t)$ by assuming as the spectral shape a log-parabola with exponential cutoff, yielding:
\begin{equation}
    \Phi_z(E,t) = \Phi(t) \left(\frac{E}{E_0}\right)^{-\Gamma(t)-\beta ln\frac{E}{E_0}} e^{-\frac{E}{E_\mathrm{cut}}} e^{-\tau_{\gamma \gamma}(E,z)}
\label{eq:spectralmodel}
\end{equation}
where $E_0$ is the reference energy, $\Phi(t)$ is the differential flux at the reference energy, $\Gamma(t)$ the photon index, $\beta$ the log-parabola curvature and $E_\mathrm{cut}$ the cutoff energy. The last factor in Equation \ref{eq:spectralmodel} describes the absorption of VHE photons in the extragalactic background light (EBL) with the optical depth $\tau(E,z)$ taken from the work of \cite{2011MNRAS.410.2556D}. The optical depth depends on the source's redshift, z, and the gamma-ray energy. The generated spectra are presented in Figure~\ref{fig:SED_BLLac} for the case of BL Lac.

\subsection{Modelling of the fast variability of AGN}

\begin{figure}[]
\begin{center}
\resizebox{.45\textwidth}{!}{\includegraphics[clip=true]{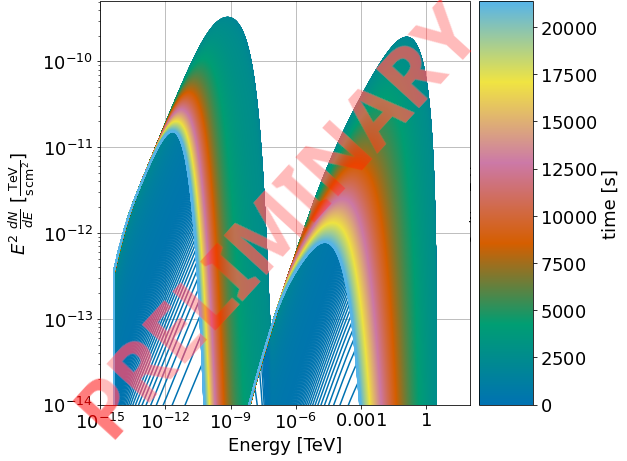}}
\end{center}
\caption{The time evolution of Mrk 421 spectrum during the flare. Colors displaying the time evolution.}
\label{fig:SED_Mrk421} 
\end{figure}

For AGN flares, we use phenomenological models inspired by observations of prominent historical flares captured with current instruments. Specifically, we focus on a model developed to reproduce the 2001 TeV flare of Mrk 421 \citep{2008ApJ...677..906F}. This model, detailed in \cite{2008ApJ...686..181F}, assumes synchrotron self-Compton (SSC) emission, where the accelerated particles are electrons. The model employs a single-zone leptonic framework, where electrons are injected with a power-law energy distribution and subsequently cool as they radiate. These electrons produce synchrotron emission and undergo inverse Compton scattering with their own synchrotron photons, resulting in the observed high-energy radiation.
The resulting time-dependent spectrum is presented in Fig.~\ref{fig:SED_Mrk421}

\section{Simulations and reconstruction of AGN observations with CTAO}
For simulating CTAO observations, we use the \textsc{CtaAgnVar}\footnote{\url{https://gitlab.cta-observatory.org/guillaume.grolleron/ctaagnvar}} tool, a pipeline specifically designed for simulating and analyzing AGN observations with CTAO. This pipeline is built on \textsc{gammapy}, the high-level analysis framework for CTAO \cite{2023A&A...678A.157D}. The \textsc{CtaAgnVar} pipeline generates an observation sequence based on an input time-dependent AGN spectral model. It incorporates CTAO’s observational constraints, including the night-time visibility of the source and the zenith angle of observation. This ensures the dynamic selection of appropriate instrument response functions (IRFs) \citep{cherenkov_telescope_array_observatory_2021_5499840}. The pipeline enables realistic simulation of gamma-ray-like events, which are subsequently fitted using analytical spectral models. Light curve (LC) reconstruction is achieved by testing multiple spectral models on the simulated data, with the best fit selected through a likelihood ratio test.

\section{Results}
\subsection{Long-term lightcurve and PSD reconstruction}
\begin{figure}[]
\begin{center}
\resizebox{0.45\textwidth}{!}{\includegraphics[clip=true]{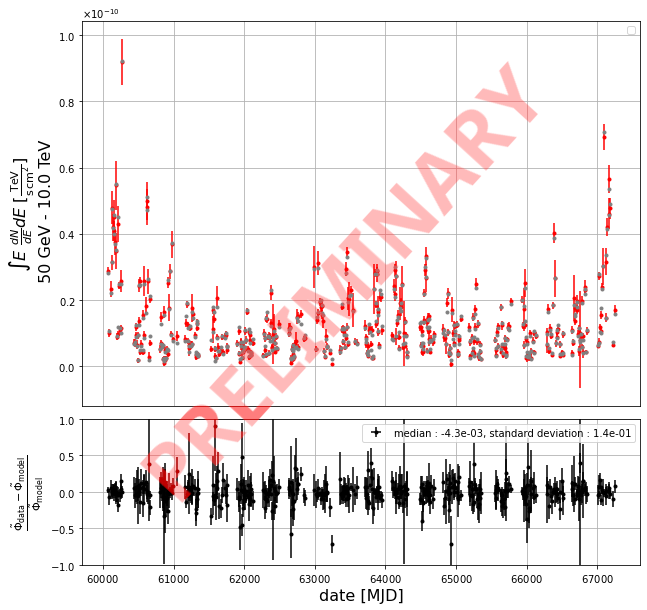}}
\end{center}
\caption{Reconstructed LC above 50 GeV and residuals computed between the simulated and reconstructed data for BL Lac for the simulation of 20 years. Gray points are the injected values and red points are the reconstructed ones.}
\label{fig:LC_BLLac} 
\end{figure}

To compute the LC, we iteratively fit each observation, starting with the simplest spectral model and progressing to more complex ones, using a likelihood ratio test for model selection. A goodness-of-fit estimator is applied to discard poorly fit spectra. The resulting LC for BL Lac, simulated over a 20-year timescale, is shown in Figure \ref{fig:LC_BLLac}. A detailed description of the LC reconstruction process is provided in \cite{2023arXiv230912157G}.

\begin{figure}[]
\begin{center}
\resizebox{0.4\textwidth}{!}{\includegraphics[clip=true]{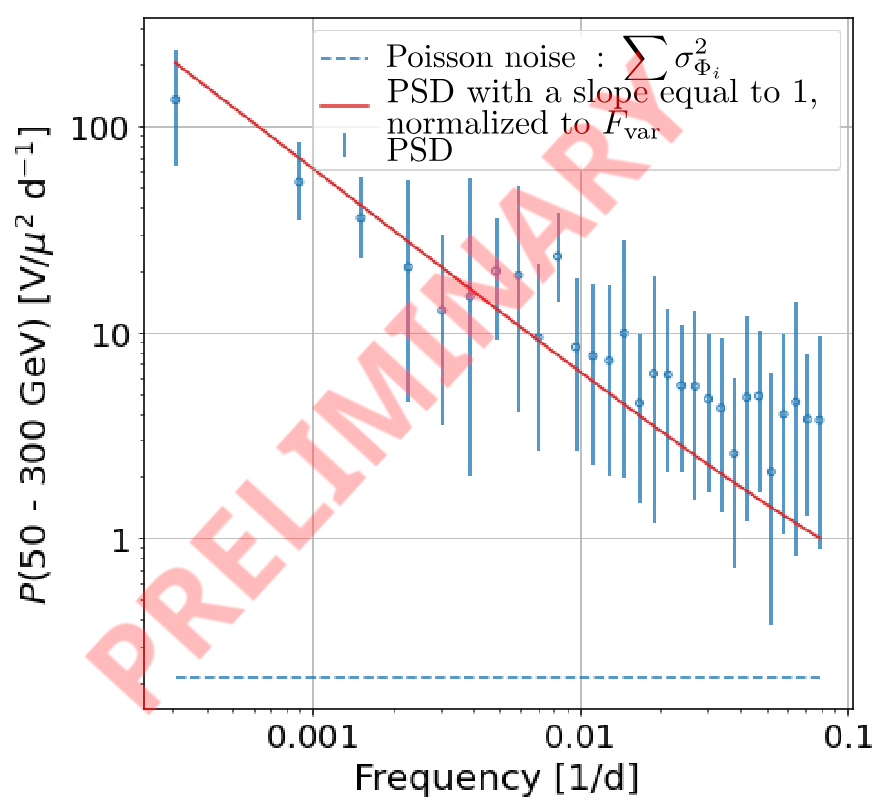}}
\end{center}
\caption{PSDs estimate (blue points) of the LC computed from simulated data for BL Lac between 50 and 300 GeV. The red line shows the injection PSD (with a floor level at high frequencies) used to simulate the input data.}
\label{fig:PSD_BLLac} 
\end{figure}

The PSD of the LC is computed following the method in \cite{2023arXiv230912157G} and is shown in Figure \ref{fig:PSD_BLLac}. Improved selection of valid LC data points results in significantly enhanced PSD reconstruction compared to the earlier work. Additionally, reconstructing the LC in narrower energy bands (e.g., above 50 GeV) yields better PSD slope accuracy than using the entire energy range. The injected PSD slope is qualitatively well-reproduced, though efforts to quantify the slope's confidence level are ongoing.

\subsection{Reconstruction of AGN flares}

\begin{figure}[]
\begin{center}
\resizebox{.42\textwidth}{!}{\includegraphics[clip=true]{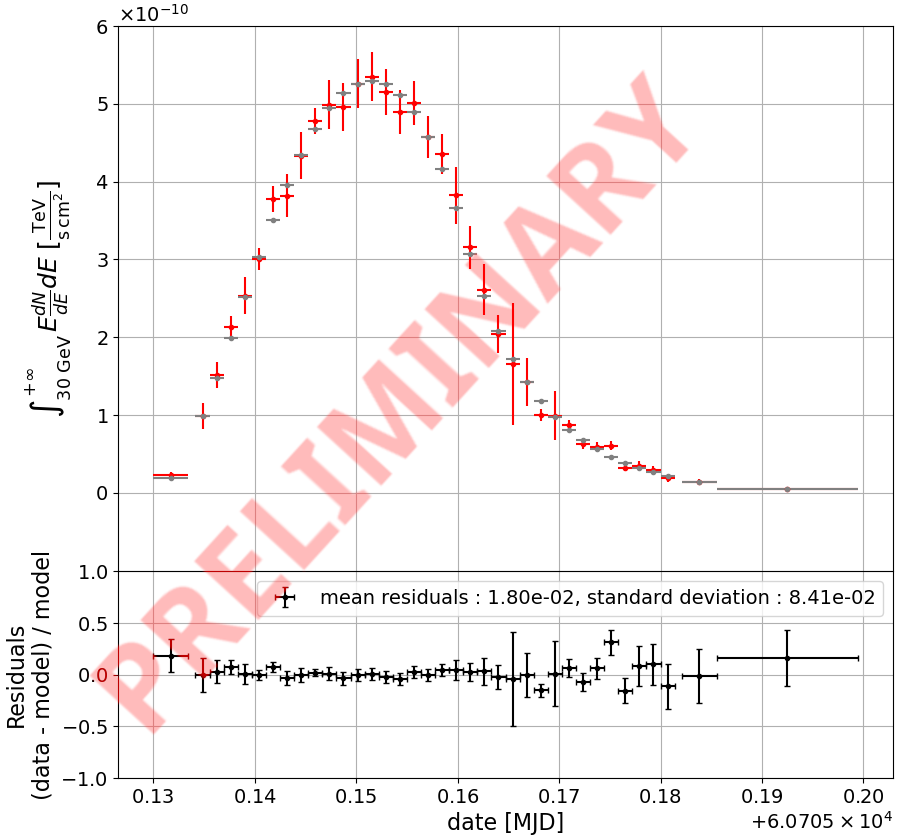}}
\end{center}
\caption{Reconstructed LC above 30 GeV and residuals computed between the simulated and reconstructed data for the Mrk 421 flare. Gray points are the injected values and red points are the reconstructed ones.}
\label{fig:LC_Mrk421} 
\end{figure}

\begin{figure}[]
\begin{center}
\resizebox{.4\textwidth}{!}{\includegraphics[clip=true]{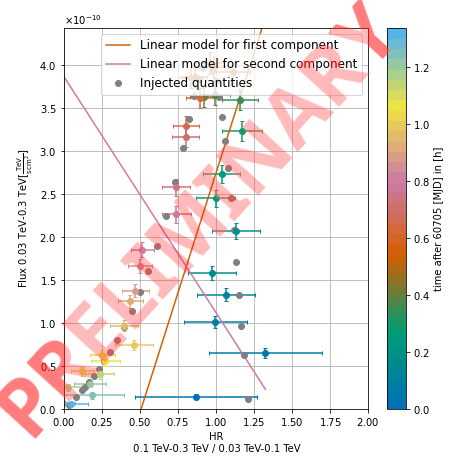}}
\end{center}
\caption{Hardness ratio diagram, with the flux between 30 and 300 GeV versus the ratio of the flux from 100 to 300 GeV, and 30 to 100 GeV. The gray points are the injected values. The color displays the time evolution. The two lines are the detected component by the PCA method.}
\label{fig:HR} 
\end{figure}

For AGN flares, the LC is computed using the same process, resulting in the LC for the Mrk 421 flare shown in Figure \ref{fig:LC_Mrk421}, calculated above 30 GeV. A variable time binning approach is employed to optimize time resolution, with bin durations set to achieve a detection significance of $5\sigma$. The reconstruction of the injected quantities shows excellent accuracy. To investigate spectral variability and distinguish between emission models, a statistical tool has been developed, leveraging hardness ratio (HR) diagrams and principal component analysis (PCA). Fully detailed in \cite{ctaagnvar}, this tool assesses the significance of hysteresis patterns in HR diagram evolution during flares. Figure \ref{fig:HR} shows the HR diagram for the Mrk 421 flare, where a qualitative hysteresis pattern is visible, quantified with a $1.5\sigma$ detection level.  Currently, the statistical method does not incorporate time information, which explains the low significance despite the clear hysteresis pattern. However, this promising method is under refinement, with improvements expected in future publications.

\section{Conclusions}
With CTAO, both AGN flares and long-term behavior will be extensively studied. For long-term behavior, this paper demonstrates that CTAO will enable precise reconstruction of the duty cycle for jetted AGN and their PSD, particularly regarding slope calculations. The next step involves quantifying CTAO's potential to detect breaks in the PSD slope, with ongoing simulations conducted using \textsc{CtaAgnVar}. For flare studies, the reconstructed LC will achieve unprecedented accuracy. Additionally, for the first time in the very-high-energy range, hysteresis in HR diagrams can be systematically investigated. To support this, we developed a statistical method for detecting hysteresis patterns. This work will contribute to an upcoming publication by the CTAO consortium.

\tiny
\noindent {\bf{Authors}}\par
J. Becerra Gonzalez$^{11}$,
J. Finke$^{12}$, 
M. Joshi$^{13}$, , 
P. Morris$^{8}$, 
M. Petropoulou$^{14}$,
A. Sarkar$^{11}$, 
P. Romano$^{15}$, 
S. Vercellone$^{15}$, 
M. Zacharias$^{16}$

\noindent {\bf{Affiliations}}\par
\tiny

$^{6}$ Departamento de Astronomía, Universidad de Chile,
Camino El Observatorio 1515, Las Condes, Santiago, Chile

$^{7}$ INFN Sezione di Padova and Università degli Studi di Padova,
Via Marzolo 8, 35131 Padova, Italy

$^{8}$ Deutsches Elektronen-Synchrotron, Platanenallee 6,
15738 Zeuthen, Germany

$^{9}$ Aalto University Metsähovi Radio Observatory,
Metsähovintie 114, 02540 Kylmälä, Finland

$^{10}$ Aalto University Department of Electronics and Nanoengineering,
PO Box 15500, 00076 Aalto, Finland

$^{11}$ Instituto de Astrofísica de Canarias and Departamento de Astrofísica, Universidad de La Laguna,
La Laguna, Tenerife, Spain

$^{12}$ Naval Research Laboratory (NRL), 4555 Overlook Ave., SW, Washington, DC 20375, United States

$^{13}$ Research Computing, Information Technology Services, Northeastern University, USA

$^{14}$ Department of Physics, National and Kapodistrian University of Athens, University Campus, Zografos,
GR 15783, Greece

$^{15}$ INAF - Osservatorio Astronomico di Brera, Via Brera 28, 20121 Milano, Italy

$^{16}$ Landessternwarte, Universität Heidelberg, Königstuhl, 69117, Heidelberg, German

\begin{acknowledgements}
\tiny
This work was conducted in the context of the CTAO Consortium. We gratefully acknowledge financial support from the agencies and organizations listed here:
\url{https://www.cta-observatory.org/consortium_acknowledgments/}.
\end{acknowledgements}

\small
\bibliographystyle{aa}
\bibliography{biblio}

@dataset{cherenkov_telescope_array_observatory_2021_5499840,
  author       = {Cherenkov Telescope Array Observatory and
                  Cherenkov Telescope Array Consortium},
  title        = {{CTAO Instrument Response Functions - prod5 version 
                   v0.1}},
  month        = sep,
  year         = 2021,
  publisher    = {Zenodo},
  version      = {v0.1},
  doi          = {10.5281/zenodo.5499840},
  url          = {https://doi.org/10.5281/zenodo.5499840}
}

@Article{1999apj...514..682e,
  author	= {{Edelson}, Rick and {Nandra}, Kirpal},
  title		= "{A Cutoff in the X-Ray Fluctuation Power Density Spectrum
		  of the Seyfert 1 Galaxy NGC 3516}",
  journal	= {\apj},
  year		= 1999,
  month		= apr,
  volume	= {514},
  number	= {2},
  pages		= {682-690},
  doi		= {10.1086/306980},
  archiveprefix	= {arXiv},
  eprint	= {astro-ph/9810481},
  primaryclass	= {astro-ph},
  adsurl	= {https://ui.adsabs.harvard.edu/abs/1999ApJ...514..682E}
}

@Article{2002mnras.332..231u,
  author	= {{Uttley}, P. and {McHardy}, I.~M. and {Papadakis}, I.~E.},
  title		= "{Measuring the broad-band power spectra of active galactic
		  nuclei with RXTE}",
  journal	= {\mnras},
  year		= 2002,
  month		= may,
  volume	= {332},
  number	= {1},
  pages		= {231-250},
  doi		= {10.1046/j.1365-8711.2002.05298.x},
  archiveprefix	= {arXiv},
  eprint	= {astro-ph/0201134},
  primaryclass	= {astro-ph},
  adsurl	= {https://ui.adsabs.harvard.edu/abs/2002MNRAS.332..231U}
}

@Article{2003mnras.345.1271v,
  author	= {{Vaughan}, S. and {Edelson}, R. and {Warwick}, R.~S. and
		  {Uttley}, P.},
  title		= "{On characterizing the variability properties of X-ray
		  light curves from active galaxies}",
  journal	= {\mnras},
  year		= 2003,
  month		= nov,
  volume	= {345},
  number	= {4},
  pages		= {1271-1284},
  doi		= {10.1046/j.1365-2966.2003.07042.x},
  archiveprefix	= {arXiv},
  eprint	= {astro-ph/0307420},
  primaryclass	= {astro-ph},
  adsurl	= {https://ui.adsabs.harvard.edu/abs/2003MNRAS.345.1271V}
}

@Article{2006natur.444..730m,
  author	= {{McHardy}, I.~M. and {Koerding}, E. and {Knigge}, C. and
		  {Uttley}, P. and {Fender}, R.~P.},
  title		= "{Active galactic nuclei as scaled-up Galactic black
		  holes}",
  journal	= {\nat},
  year		= 2006,
  month		= dec,
  volume	= {444},
  number	= {7120},
  pages		= {730-732},
  doi		= {10.1038/nature05389},
  archiveprefix	= {arXiv},
  eprint	= {astro-ph/0612273},
  primaryclass	= {astro-ph},
  adsurl	= {https://ui.adsabs.harvard.edu/abs/2006Natur.444..730M}
}

@Article{2008apj...677..906f,
  author	= {{Fossati}, G. and {Buckley}, J.~H. and {Bond}, I.~H. and
		  {Bradbury}, S.~M. and {Carter-Lewis}, D.~A. and {Chow},
		  Y.~C.~K. and {Cui}, W. and {Falcone}, A.~D. and {Finley},
		  J.~P. and {Gaidos}, J.~A. and {Grube}, J. and {Holder}, J.
		  and {Horan}, D. and {Horns}, D. and {Jordan}, M.~M. and
		  {Kieda}, D.~B. and {Kildea}, J. and {Krawczynski}, H. and
		  {Krennrich}, F. and {Lang}, M.~J. and {LeBohec}, S. and
		  {Lee}, K. and {Moriarty}, P. and {Ong}, R.~A. and {Petry},
		  D. and {Quinn}, J. and {Sembroski}, G.~H. and {Wakely},
		  S.~P. and {Weekes}, T.~C.},
  title		= "{Multiwavelength Observations of Markarian 421 in 2001
		  March: An Unprecedented View on the X-Ray/TeV Correlated
		  Variability}",
  journal	= {\apj},
  year		= 2008,
  month		= apr,
  volume	= {677},
  number	= {2},
  pages		= {906-925},
  doi		= {10.1086/527311},
  archiveprefix	= {arXiv},
  eprint	= {0710.4138},
  primaryclass	= {astro-ph},
  adsurl	= {https://ui.adsabs.harvard.edu/abs/2008ApJ...677..906F}
}

@Article{2008apj...686..181f,
  author	= {{Finke}, Justin D. and {Dermer}, Charles D. and
		  {B{\"o}ttcher}, Markus},
  title		= "{Synchrotron Self-Compton Analysis of TeV X-Ray-Selected
		  BL Lacertae Objects}",
  journal	= {\apj},
  year		= 2008,
  month		= oct,
  volume	= {686},
  number	= {1},
  pages		= {181-194},
  doi		= {10.1086/590900},
  archiveprefix	= {arXiv},
  eprint	= {0802.1529},
  primaryclass	= {astro-ph},
  adsurl	= {https://ui.adsabs.harvard.edu/abs/2008ApJ...686..181F}
}

@Article{2010a&a...520a..83h,
  author	= {{H.~E.~S.~S. Collaboration} and {Abramowski}, A. and
		  {Acero}, F. and {Aharonian}, F. and {Akhperjanian}, A.~G.
		  and {Anton}, G. and {Barres de Almeida}, U. and
		  {Bazer-Bachi}, A.~R. and {Becherini}, Y. Behera, B. and
		  {Benbow}, W. and {Bernl{\"o}hr}, K. and {Bochow}, A. and
		  {Boisson}, C. and {Bolmont}, J. and {Borrel}, V. and
		  {Brucker}, J. and {Brun}, F. and {Brun}, P. and
		  {B{\"u}hler}, R. and {Bulik}, T. and {B{\"u}sching}, I. and
		  {Boutelier}, T. and {Chadwick}, P.~M. and {Charbonnier}, A.
		  and {Chaves}, R.~C.~G. and {Cheesebrough}, A. and
		  {Chounet}, L. -M. and {Clapson}, A.~C. and {Coignet}, G.
		  and {Conrad}, J. and {Costamante}, L. and {Dalton}, M. and
		  {Daniel}, M.~K. and {Davids}, I.~D. and {Degrange}, B. and
		  {Deil}, C. and {Dickinson}, H.~J. and {Djannati-Ata{\"\i}},
		  A. and {Domainko}, W. and {O'C. Drury}, L. and {Dubois}, F.
		  and {Dubus}, G. and {Dyks}, J. and {Dyrda}, M. and
		  {Egberts}, K. and {Eger}, P. and {Espigat}, P. and
		  {Fallon}, L. and {Farnier}, C. and {Fegan}, S. and
		  {Feinstein}, F. and {Fernandes}, M.~V. and {Fiasson}, A.
		  and {F{\"o}rster}, A. and {Fontaine}, G. and
		  {F{\"u}{\ss}ling}, M. and {Gabici}, S. and {Gallant}, Y.~A.
		  and {G{\'e}rard}, L. and {Gerbig}, D. and {Giebels}, B. and
		  {Glicenstein}, J.~F. and {Gl{\"u}ck}, B. and {Goret}, P.
		  and {G{\"o}ring}, D. and {Hampf}, D. and {Hauser}, M. and
		  {Heinz}, S. and {Heinzelmann}, G. and {Henri}, G. and
		  {Hermann}, G. and {Hinton}, J.~A. and {Hoffmann}, A. and
		  {Hofmann}, W. and {Hofverberg}, P. and {Holleran}, M. and
		  {Hoppe}, S. and {Horns}, D. and {Jacholkowska}, A. and {de
		  Jager}, O.~C. and {Jahn}, C. and {Jung}, I. and
		  {Katarzy{\'n}ski}, K. and {Katz}, U. and {Kaufmann}, S. and
		  {Kerschhaggl}, M. and {Khangulyan}, D. and {Kh{\'e}lifi},
		  B. and {Keogh}, D. and {Klochkov}, D. and {Klu{\'z}niak},
		  W. and {Kneiske}, T. and {Komin}, Nu. and {Kosack}, K. and
		  {Kossakowski}, R. and {Lamanna}, G. and {Lenain}, J.-P. and
		  {Lohse}, T. and {Lu}, C. -C. and {Marandon}, V. and
		  {Marcowith}, A. and {Masbou}, J. and {Maurin}, D. and
		  {McComb}, T.~J.~L. and {Medina}, M.~C. and {M{\'e}hault},
		  J. and {Moderski}, R. and {Moulin}, E. and {Naumann-Godo},
		  M. and {de Naurois}, M. and {Nedbal}, D. and {Nekrassov},
		  D. and {Nguyen}, N. and {Nicholas}, B. and {Niemiec}, J.
		  and {Nolan}, S.~J. and {Ohm}, S. and {Olive}, J. -F. and
		  {de O{\~n}a Wilhelmi}, E. and {Opitz}, B. and {Orford},
		  K.~J. and {Ostrowski}, M. and {Panter}, M. and {Paz
		  Arribas}, M. and {Pedaletti}, G. and {Pelletier}, G. and
		  {Petrucci}, P. -O. and {Pita}, S. and {P{\"u}hlhofer}, G.
		  and {Punch}, M. and {Quirrenbach}, A. and {Raubenheimer},
		  B.~C. and {Raue}, M. and {Rayner}, S.~M. and {Reimer}, O.
		  and {Renaud}, M. and {de los Reyes}, R. and {Rieger}, F.
		  and {Ripken}, J. and {Rob}, L. and {Rosier-Lees}, S. and
		  {Rowell}, G. and {Rudak}, B. and {Rulten}, C.~B. and
		  {Ruppel}, J. and {Ryde}, F. and {Sahakian}, V. and
		  {Santangelo}, A. and {Schlickeiser}, R. and {Sch{\"o}ck},
		  F.~M. and {Sch{\"o}nwald}, A. and {Schwanke}, U. and
		  {Schwarzburg}, S. and {Schwemmer}, S. and {Shalchi}, A. and
		  {Sushch}, I. and {Sikora}, M. and {Skilton}, J.~L. and
		  {Sol}, H. and {Stawarz}, {\L}. and {Steenkamp}, R. and
		  {Stegmann}, C. and {Stinzing}, F. and {Superina}, G. and
		  {Szostek}, A. and {Tam}, P.~H. and {Tavernet}, J. -P. and
		  {Terrier}, R. and {Tibolla}, O. and {Tluczykont}, M. and
		  {Valerius}, K. and {van Eldik}, C. and {Vasileiadis}, G.
		  and {Venter}, C. and {Venter}, L. and {Vialle}, J.~P. and
		  {Viana}, A. and {Vincent}, P. and {Vivier}, M. and
		  {V{\"o}lk}, H.~J. and {Volpe}, F. and {Vorobiov}, S. and
		  {Wagner}, S.~J. and {Ward}, M. and {Zdziarski}, A.~A. and
		  {Zech}, A. and {Zechlin}, H.~S.},
  title		= "{VHE {\ensuremath{\gamma}}-ray emission of PKS 2155-304:
		  spectral and temporal variability}",
  journal	= {\aap},
  year		= 2010,
  month		= sep,
  volume	= {520},
  eid		= {A83},
  pages		= {A83},
  doi		= {10.1051/0004-6361/201014484},
  archiveprefix	= {arXiv},
  eprint	= {1005.3702},
  primaryclass	= {astro-ph.HE},
  adsurl	= {https://ui.adsabs.harvard.edu/abs/2010A&A...520A..83H}
}

@Article{2011mnras.410.2556d,
  author	= {{Dom{\'\i}nguez}, A. and {Primack}, J.~R. and {Rosario},
		  D.~J. and {Prada}, F. and {Gilmore}, R.~C. and {Faber},
		  S.~M. and {Koo}, D.~C. and {Somerville}, R.~S. and
		  {P{\'e}rez-Torres}, M.~A. and {P{\'e}rez-Gonz{\'a}lez}, P.
		  and {Huang}, J. -S. and {Davis}, M. and {Guhathakurta}, P.
		  and {Barmby}, P. and {Conselice}, C.~J. and {Lozano}, M.
		  and {Newman}, J.~A. and {Cooper}, M.~C.},
  title		= "{Extragalactic background light inferred from AEGIS
		  galaxy-SED-type fractions}",
  journal	= {\mnras},
  year		= 2011,
  month		= feb,
  volume	= {410},
  number	= {4},
  pages		= {2556-2578},
  doi		= {10.1111/j.1365-2966.2010.17631.x},
  archiveprefix	= {arXiv},
  eprint	= {1007.1459},
  primaryclass	= {astro-ph.CO},
  adsurl	= {https://ui.adsabs.harvard.edu/abs/2011MNRAS.410.2556D}
}

@Article{2013mnras.433..907e,
  author	= {{Emmanoulopoulos}, D. and {McHardy}, I.~M. and
		  {Papadakis}, I.~E.},
  title		= "{Generating artificial light curves: revisited and
		  updated}",
  journal	= {\mnras},
  year		= 2013,
  month		= aug,
  volume	= {433},
  number	= {2},
  pages		= {907-927},
  doi		= {10.1093/mnras/stt764},
  archiveprefix	= {arXiv},
  eprint	= {1305.0304},
  primaryclass	= {astro-ph.IM},
  adsurl	= {https://ui.adsabs.harvard.edu/abs/2013MNRAS.433..907E}
}

@Article{2014mnras.444.1077k,
  author	= {{Kapanadze}, B. and {Romano}, P. and {Vercellone}, S. and
		  {Kapanadze}, S.},
  title		= "{The X-ray behaviour of the high-energy peaked BL Lacertae
		  source PKS 2155-304 in the 0.3-10 keV band}",
  journal	= {\mnras},
  year		= 2014,
  month		= oct,
  volume	= {444},
  number	= {2},
  pages		= {1077-1094},
  doi		= {10.1093/mnras/stu1504},
  adsurl	= {https://ui.adsabs.harvard.edu/abs/2014MNRAS.444.1077K}
}

@Article{2019ara&a..57..467b,
  author	= {{Blandford}, Roger and {Meier}, David and {Readhead},
		  Anthony},
  title		= "{Relativistic Jets from Active Galactic Nuclei}",
  journal	= {ARA\&A},
  year		= 2019,
  month		= aug,
  volume	= {57},
  pages		= {467-509},
  doi		= {10.1146/annurev-astro-081817-051948},
  archiveprefix	= {arXiv},
  eprint	= {1812.06025},
  primaryclass	= {astro-ph.HE},
  adsurl	= {https://ui.adsabs.harvard.edu/abs/2019ARA&A..57..467B}
}

@Article{2019galax...7...28r,
  author	= {{Rieger}, Frank},
  title		= "{Gamma-Ray Astrophysics in the Time Domain}",
  journal	= {Galaxies},
  year		= 2019,
  month		= jan,
  volume	= {7},
  number	= {1},
  pages		= {28},
  doi		= {10.3390/galaxies7010028},
  archiveprefix	= {arXiv},
  eprint	= {1901.10216},
  primaryclass	= {astro-ph.HE},
  adsurl	= {https://ui.adsabs.harvard.edu/abs/2019Galax...7...28R}
}

@Article{2019newar..8701541h,
  author	= {{Hovatta}, Talvikki and {Lindfors}, Elina},
  title		= "{Relativistic Jets of Blazars}",
  journal	= {\nar},
  year		= 2019,
  month		= dec,
  volume	= {87},
  eid		= {101541},
  pages		= {101541},
  doi		= {10.1016/j.newar.2020.101541},
  archiveprefix	= {arXiv},
  eprint	= {2003.06322},
  primaryclass	= {astro-ph.HE},
  adsurl	= {https://ui.adsabs.harvard.edu/abs/2019NewAR..8701541H}
}

@Article{2020galax...8...72c,
  author	= {{Cerruti}, Matteo},
  title		= "{Leptonic and Hadronic Radiative Processes in
		  Supermassive-Black-Hole Jets}",
  journal	= {Galaxies},
  year		= 2020,
  month		= oct,
  volume	= {8},
  number	= {4},
  eid		= {72},
  pages		= {72},
  doi		= {10.3390/galaxies8040072},
  archiveprefix	= {arXiv},
  eprint	= {2012.13302},
  primaryclass	= {astro-ph.HE},
  adsurl	= {https://ui.adsabs.harvard.edu/abs/2020Galax...8...72C}
}

@Article{2023a&a...678a.157d,
  author	= {{Donath}, Axel and {Terrier}, R{\'e}gis and {Remy},
		  Quentin and {Sinha}, Atreyee and {Nigro}, Cosimo and
		  {Pintore}, Fabio and {Kh{\'e}lifi}, Bruno and
		  {Olivera-Nieto}, Laura and {Ruiz}, Jose Enrique and
		  {Br{\"u}gge}, Kai and {Linhoff}, Maximilian and
		  {Contreras}, Jose Luis and {Acero}, Fabio and
		  {Aguasca-Cabot}, Arnau and {Berge}, David and
		  {Bhattacharjee}, Pooja and {Buchner}, Johannes and
		  {Boisson}, Catherine and {Carreto Fidalgo}, David and
		  {Chen}, Andrew and {de Bony de Lavergne}, Mathieu and {de
		  Miranda Cardoso}, Jos{\'e} Vinicius and {Deil}, Christoph
		  and {F{\"u}{\ss}ling}, Matthias and {Funk}, Stefan and
		  {Giunti}, Luca and {Hinton}, Jim and {Jouvin}, L{\'e}a and
		  {King}, Johannes and {Lefaucheur}, Julien and
		  {Lemoine-Goumard}, Marianne and {Lenain}, Jean-Philippe and
		  {L{\'o}pez-Coto}, Rub{\'e}n and {Mohrmann}, Lars and
		  {Morcuende}, Daniel and {Panny}, Sebastian and {Regeard},
		  Maxime and {Saha}, Lab and {Siejkowski}, Hubert and
		  {Siemiginowska}, Aneta and {Sip{\H{o}}cz}, Brigitta M. and
		  {Unbehaun}, Tim and {van Eldik}, Christopher and
		  {Vuillaume}, Thomas and {Zanin}, Roberta},
  title		= "{Gammapy: A Python package for gamma-ray astronomy}",
  journal	= {\aap},
  year		= 2023,
  month		= oct,
  volume	= {678},
  eid		= {A157},
  pages		= {A157},
  doi		= {10.1051/0004-6361/202346488},
  archiveprefix	= {arXiv},
  eprint	= {2308.13584},
  primaryclass	= {astro-ph.IM},
  adsurl	= {https://ui.adsabs.harvard.edu/abs/2023A&A...678A.157D}
}

@Article{2023arxiv230400835g,
  author	= {{Gr{\'e}aux}, Lucas and {Biteau}, Jonathan and {Hassan},
		  Tarek and {Hervet}, Olivier and {Nievas Rosillo}, Mireia
		  and {Williams}, David A.},
  title		= "{STeVECat, the Spectral TeV Extragalactic Catalog}",
  journal	= {arXiv e-prints},
  year		= 2023,
  month		= apr,
  eid		= {arXiv:2304.00835},
  pages		= {arXiv:2304.00835},
  doi		= {10.48550/arXiv.2304.00835},
  archiveprefix	= {arXiv},
  eprint	= {2304.00835},
  primaryclass	= {astro-ph.HE},
  adsurl	= {https://ui.adsabs.harvard.edu/abs/2023arXiv230400835G}
}

@Article{2023arxiv230414208c,
  author	= {{Cangemi}, F. and {Hovatta}, T. and {Lindfors}, E. and
		  {Cerruti}, M. and {Becerra-Gonzalez}, J. and {Biteau}, J.
		  and {Boisson}, C. and {B{\"o}ttcher}, M. and {de Gouveia
		  Dal Pino}, E. and {Dorner}, D. and {Grolleron}, G. and
		  {Lenain}, J. -P. and {Manganaro}, M. and {Max-Moerbeck}, W.
		  and {Morris}, P. and {Nilsson}, K. and {Passos Reis}, L.
		  and {Romano}, P. and {Sergijenko}, O. and {Tavecchio}, F.
		  and {Vercellone}, S. and {Wagner}, S. and {Zacharias}, M.},
  title		= "{Probing AGN variability with the Cherenkov Telescope
		  Array}",
  journal	= {arXiv e-prints},
  year		= 2023,
  month		= apr,
  eid		= {arXiv:2304.14208},
  pages		= {arXiv:2304.14208},
  doi		= {10.48550/arXiv.2304.14208},
  archiveprefix	= {arXiv},
  eprint	= {2304.14208},
  primaryclass	= {astro-ph.HE},
  adsurl	= {https://ui.adsabs.harvard.edu/abs/2023arXiv230414208C}
}

@Article{ctaagnvar,
  author	= {{Grolleron}, Guillaume and {Lenain}, Jean-Philippe},
  title		= "{\textsc{CtaAgnVar} : a Pipeline for Variability Studies
		  of blazars with the Cherenkov Telescope Array}",
  year		= "2025",
  journal	= "in prep."
}

@InBook{	  inbook,
  author	= {Zech, A. and Mazin, Daniel and Biteau, Jonathan and
		  Daniel, Michael and Hassan, Tarek and Lindfors, E. and
		  Meyer, Manuel},
  year		= {2019},
  month		= {02},
  pages		= {231-272},
  title		= {KSP: Active Galactic Nuclei},
  isbn		= {978-981-327-008-4},
  doi		= {10.1142/9789813270091_0012}
}

\end{document}